%% file: PAM2020.tex
\documentclass[runningheads]{llncs}

\usepackage{xcolor}
\usepackage[tableposition=top]{caption}
\usepackage{ftnxtra}
\usepackage{color, colortbl}
\usepackage{booktabs, array, dcolumn}
\usepackage{tikz}
\usepackage{graphicx}
\usepackage{pgf-umlsd}
\usepackage{multirow}
\usepackage{subfig}
\usepackage{pgfplotstable}
\usepackage{amsfonts}
\usepackage{relsize}
\usepackage{varwidth}
\usepackage{url}
\usepackage{amsmath}
\usepackage{amssymb}
\usepackage[normalem]{ulem}
\usepackage{flushend}
\usepackage[flushleft]{threeparttable}
\usepackage[ruled]{algorithm2e}
\usepackage{makecell}
\usepackage{hyperref}

\begin{document}

\title{Characterizing and Detecting Money Laundering Activities on the Bitcoin Network}

\author{Yining Hu\inst{1,2}\orcidID{0000-0002-6233-743X} \and
Suranga Seneviratne \inst{3}\orcidID{0000-0002-5485-5595} \and
Kanchana Thilakarathna \inst{3}\orcidID{0000-0003-4332-0082} \and
Kensuke Fukuda \inst{4}\orcidID{0000-0001-8372-2807} \and
Aruna Seneviratne \inst{2}\orcidID{0000-0001-6894-7987}}
\authorrunning{Y. Hu et al.}
\institute{Data61-CSIRO, Sydney, Australia\\
\email{Yining.Hu@data61.csiro.au} \and
University of New South Wales, Sydney, Australia\\
\email{a.seneviratne@unsw.edu.au} \and
University of Sydney, Sydney, Australia\\
\email{\{firstname.lastname\}@sydney.edu.au} \and 
National Institute of Informatics, Tokyo, Japan\\
\email{kensuke@nii.ac.jp}}



 


\maketitle

\begin{abstract}
Bitcoin is by far the most popular crypto-currency solution enabling peer-to-peer payments. Despite some studies highlighting the network does not provide full anonymity, it is still being heavily used for a wide variety of dubious financial activities such as money laundering, ponzi schemes, and ransom-ware payments. In this paper, we explore the landscape of potential money laundering activities occurring across the Bitcoin network. Using data collected over three years, we create transaction graphs and provide an in-depth analysis on various graph characteristics to differentiate money laundering transactions from regular transactions. 
We found that the main difference between laundering and regular transactions lies in their output values and neighbourhood information. 
Then, we propose and evaluate a set of classifiers based on four types of graph features: immediate neighbours, curated features, deepwalk embeddings, and node2vec embeddings to classify money laundering and regular transactions.
Results show that the node2vec-based classifier outperforms other classifiers in binary classification reaching an average accuracy of 92.29\% and an F1-measure of 0.93 and high robustness over a 2.5-year time span.
Finally, we demonstrate how effective our classifiers are in discovering unknown laundering services. The classifier performance dropped compared to binary classification, however, the prediction can be improved with simple ensemble techniques for some services. 

\keywords{Bitcoin, Transaction Graph, Money Laundering}
\end{abstract}

\input{./introduction}
\input{./related}

\input{./data}
\input{./graph_characterization}
\input{./service_combined}
\input{./service_separate}
\input{./discussion_conclusion}
\input{./appendix}

\bibliographystyle{splncs04}
\bibliography{bibliography}
	
\end{document}

%% file: introduction.tex
\section{Introduction}
\label{sec:intro}
The first successful peer-to-peer (P2P) financial system, Bitcoin~\cite{nakamoto2008bitcoin}, has evolved over the years into a complex financial ecosystem with a large number of users, different transaction types, and various support services. As of now, Bitcoin has the highest market share of ~50\% among crypto-currencies~\cite{bitcoin_share} and performs over 300k transactions per day~\cite{blockchain_info}. 

Driven by the pseudonymity provided by the network, many cyber-criminals and hackers have started using Bitcoin for illegal activities. For example, \emph{Silk Road}~\cite{SilkRoad}, an online market place for illegal goods and services, accepted bitcoins as their payment method to hide the identities of the sellers and buyers. More recently, ransom-ware attacks \emph{WannaCry} and \emph{Petaya}~\cite{ransomeware_news} also accepted Bitcoin payments from victimized computer owners to unlock their machines. Other activities that involve the misuse of Bitcoin include money laundering via mixing to hide the origin of illegally obtained money~\cite{moser2013inquiry,de2017analysis}, and ponzi scheme, a pyramid scheme that pays old users with investments from new users~\cite{bartoletti2018data}.


In this paper, we investigate money laundering activities, one of the main misuses of Bitcoin~\cite{moser2013inquiry}. While there are public websites such as Blockchain Info~\cite{blockchain_info}, BitcoinWhosWho~\cite{btc_whoswho}, WalletExplorer.com~\cite{wallet_explorer} that collect de-anonymized Bitcoin services and tag those involved in money laundering, new money laundering services emerge frequently due to the unregulated and P2P nature of Bitcoin. 
Although several efforts have been made to understand and detect laundering activities~\cite{bartoletti2018data,de2017analysis,weber2019anti}, existing studies have not looked at the graph properties of laundering and regular transactions in detail, they also have not fully explored the potential of automatically created node embeddings using techniques such as deepwalk~\cite{perozzi2014deepwalk} and node2vec~\cite{grover2016node2vec}.
In this paper, we first explore money laundering transactions on the Bitcoin network from a graph theoretic perspective and compare their characteristics to regular transactions. We find that although some manually extracted statistical and network features follow different distributions for laundering and regular transactions, they are not sufficiently effective in detecting laundering transactions. We show that random-walk based graph representation learning algorithms--deepwalk~\cite{perozzi2014deepwalk} and node2vec~\cite{grover2016node2vec} significantly outperform manually created features for this task. This paper makes the following contributions:
\begin{itemize}
    \item With Bitcoin data collected over three years, we characterize graph properties of money laundering transactions and highlight their differences in comparison to regular transactions. 
    \item We show that laundering transactions are distinguishable from regular transactions in several statistical and network features including in-degree/out-degree ratio, sum/mean/standard deviation of output values, and number of weakly connected components--the size of the subgraph a transaction belongs to. Nonetheless, these metrics are not effective for the binary classification of money laundering and regular transactions.
    \item We show a node2vec-based classifier achieves the highest performance in classifying laundering and regular transactions. We also show the robustness of the classifier by applying it to randomly selected weeks across a large timescale of two and a half years and show that results remain consistent. 
    \item Finally, we demonstrate the performance of our classifiers in detecting unknown money laundering transactions. The classifier performance decreases compared to the binary classification, but can be improved with simple ensemble techniques.
\end{itemize}

The remainder of the paper is organized as follows. Section~\ref{sec:related} discusses related work that motivated our research. Section~\ref{sec:data} presents details of our data collection, ground truth labelling and how we created the Bitcoin transaction graphs. Section~\ref{sec:graph_characterization} characterizes the properties of money laundering transactions in comparison to regular transactions. Section~\ref{sec:classification} presents the classifier design and classification results, followed by Section~\ref{sec:prediction} which presents the new money laundering service discovery results. Section~\ref{sec:discussion_conclusion} concludes the paper.

%% file: related.tex
\section{Related Work} \label{sec:related}
\paragraph{\bf Bitcoin Network Characterization} 
Ron and Shamir~\cite{ron2013quantitative} found a significant variance in the distribution of various Bitcoin addresses, accumulated balance and number of transactions per user providing empirical evidence that a limited number of Bitcoin entities control the majority of addresses, transactions and bitcoins. Lischke et al.~\cite{lischke2016analyzing} conducted a study on the Bitcoin transaction graph during its first four years. Authors observed the distribution of several graph metrics such as in-degree, out-degree and clustering coefficient of the entire transaction graph. 
They also analyzed the economic and network aspects of multiple major Bitcoin businesses and markets including \emph{SatoshiDice} and \emph{Mt.Gox}. Similar studies include~\cite{parino2018analysis,somin2018social}. In this paper, we analyse the graph characteristics of potential laundering and regular transactions separately with the objective of coming up with a transaction classifier.


\paragraph{\bf Address De-anonymization} 
Multiple studies explored the possibility of de-anonymizing Bitcoin addresses~\cite{reid2011analysis,meiklejohn2013fistful,ermilov2017automatic}. Reid et al.~\cite{reid2011analysis} found external information, such as user registration details and voluntary disclosure of public-keys, can be used to link Bitcoin addresses to real-life users. Meiklejohn et al.~\cite{meiklejohn2013fistful} proposed two address clustering heuristics to aggregate Bitcoin transactions: i) addresses associated with the input UTXOs of a transaction belonging to the same user and ii) the change address that is created when the sum of input UTXOs exceeds the amount to pay also belongs to the sender.
Using these heuristics, a number of services and their Bitcoin addresses have been identified in the literature~\cite{de2017analysis,bartoletti2018data} and online forums~\cite{bitcointalk,wallet_explorer,btc_whoswho,blockchain_info}. 
We establish our ground truth labelling based on these identified services and their addresses.



\paragraph{\bf Service Analysis and Anomaly Detection}
Driven by its pseudonymity, there also exists a number of dubious and potentially illegal services in the Bitcoin network. Moser et al.~\cite{moser2013inquiry} found services such as \emph{Bitcoin Fog}, that hide transaction origins by withholding multiple small inputs and bundling them into a smaller number of larger outputs. Ferrin et al.~\cite{ferrin2015preliminary} later discovered a common pattern of transaction mixing which is to form a ``mixing cloud" that contains multiple interconnected ``joint transactions", which result from layering multiple transactions into a single larger transaction. 


More recently, machine learning methods are being applied to detect potentially illegal Bitcoin activities. Early studies often relied on the metadata and the temporal features of addresses or transactions. Pham et al.~\cite{pham2016anomaly} performed k-means clustering on both the transaction graph and user graph and applied outlier detection to find suspicious transactions and users, however they were only able to detect one out of 30 known cases of theft. Bartoletti et al.~\cite{bartoletti2018data} applied several supervised learning techniques to detect Bitcoin addresses associated with ponzi schemes using temporal features such as total address lifetime and active days.
Weber et al.~\cite{weber2019anti} recently presented a study on detecting Bitcoin laundering transactions using network features and node embeddings. The authors created 49 independent graphs over a total period of 2 weeks, using the first 34 graphs for training and the rest for testing and achieved an F1-measure of over 0.7 with Graph Convolutional Networks (GCN) and EvolveGCN. They also published a dataset containing the extracted network features, an anatomized edge list and labels~\cite{elliptic_data}. 
However, no information was provided on the ground truth labelling process as well as the exact features the authors used, making it hard to apply the solution in other datasets.
While our work is complementary to this study, we analyse data in a much larger time span of over three years, evaluate quantitatively different types of metadata and graph features, and address more realistic and challenging scenarios of operating on larger graphs and discovering new money laundering instances.

%% file: data.tex
\section{Data}
\label{sec:data}
We next describe how the transaction graphs are built, our data collection process, and the methodology of establishing the ground truth.

\subsection{Bitcoin Transaction Graph}
\label{subsec:basics}
\begin{figure}[!htbp]
	\centering
	\includegraphics[width=.5\columnwidth]{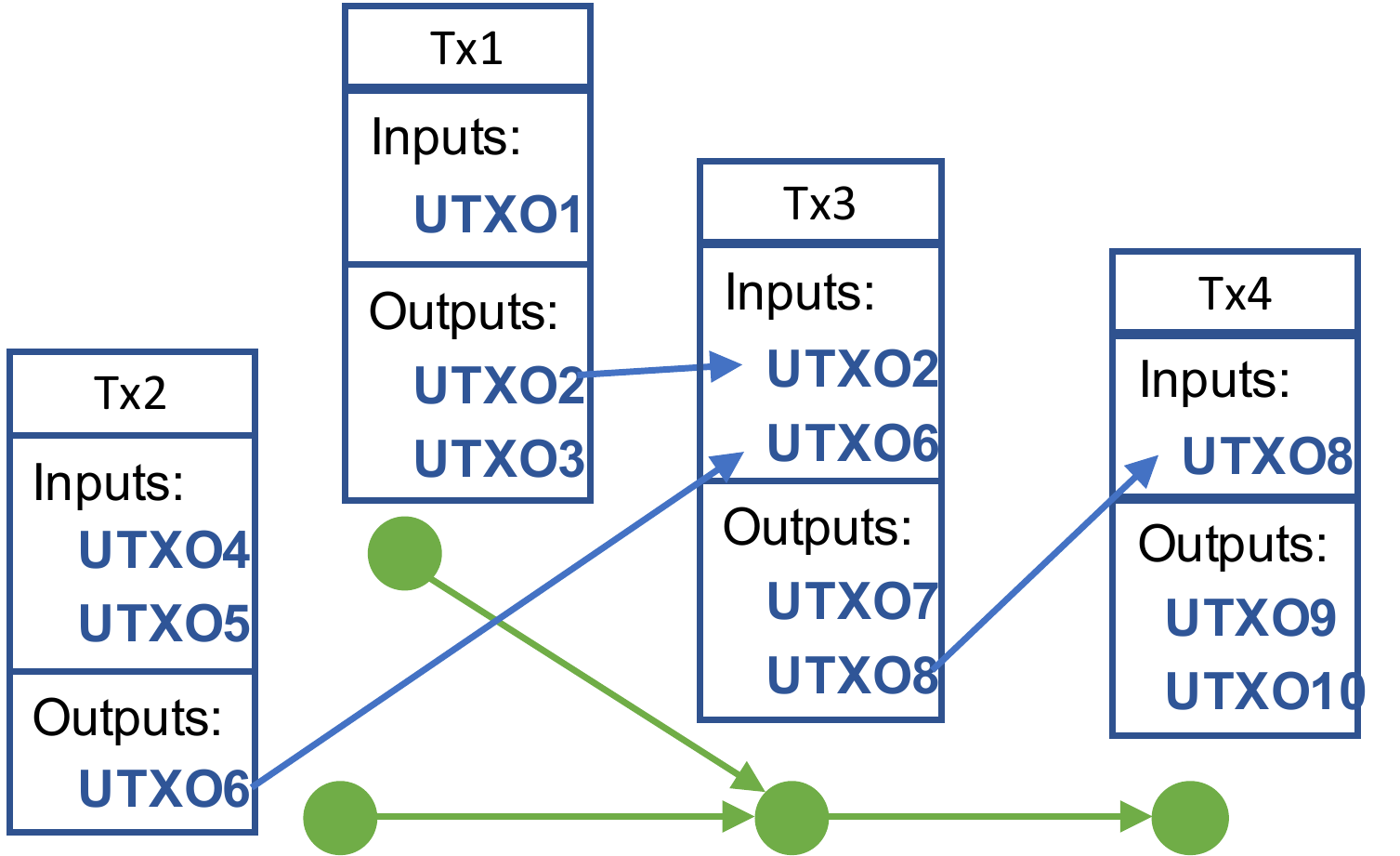} 
	\caption{A simple Bitcoin transaction chain.}
	\label{fig:transaction_illustration}
\end{figure}

A \emph{Bitcoin transaction}, identified by a unique id, comprises a list of input and output \emph{unspent transaction outputs}, or \emph{UTXOs}. UTXOs are indivisible chunks of bitcoins attached to specific owners. A Bitcoin transaction consumes UTXOs by unlocking them with the sender's signature, and creates new UTXOs designated to recipients. This is how bitcoins are transferred among users, i.e., by creating chains of transactions as shown in Figure~\ref{fig:transaction_illustration}. Each node represents a transaction and a directed edge between two nodes exists if an output UTXO of a transaction becomes an input UTXO of a succeeding transaction. For example, an edge is created from transaction \emph{Tx3} to transaction \emph{Tx4} as UTXO8 serves as an output for \emph{Tx3} and an input for \emph{Tx4}. We utilize these transaction chains to create transaction graphs. We did not explore user (address) graphs in this paper, as new addresses can be easily created and manipulated.

\vspace{-2mm}
\vspace{-2mm}



\subsection{Data Collection and Labelling}
\label{subsec:data_collect_label}
\subsubsection{Data Collection}
We ran a Bitcoin Core client under the version bitcoin-0.15.0~\cite{bitcoin_core} to collect block data, and parsed the block data with a simple parser~\cite{parser} to obtain transaction information. For each transaction, we extracted its timestamp, previous transactions (cf. Part~\ref{subsec:basics}), number and value of input and output UTXOs. Our dataset was collected between 07/2014 and 05/2017.


\subsubsection{Ground Truth Labelling}
We identify several laundering and regular services and their addresses. We consider all transactions associated with these addresses as our labelled ground truth.

\begin{figure}[!htbp]
	\centering
	\includegraphics[width=.85\columnwidth]{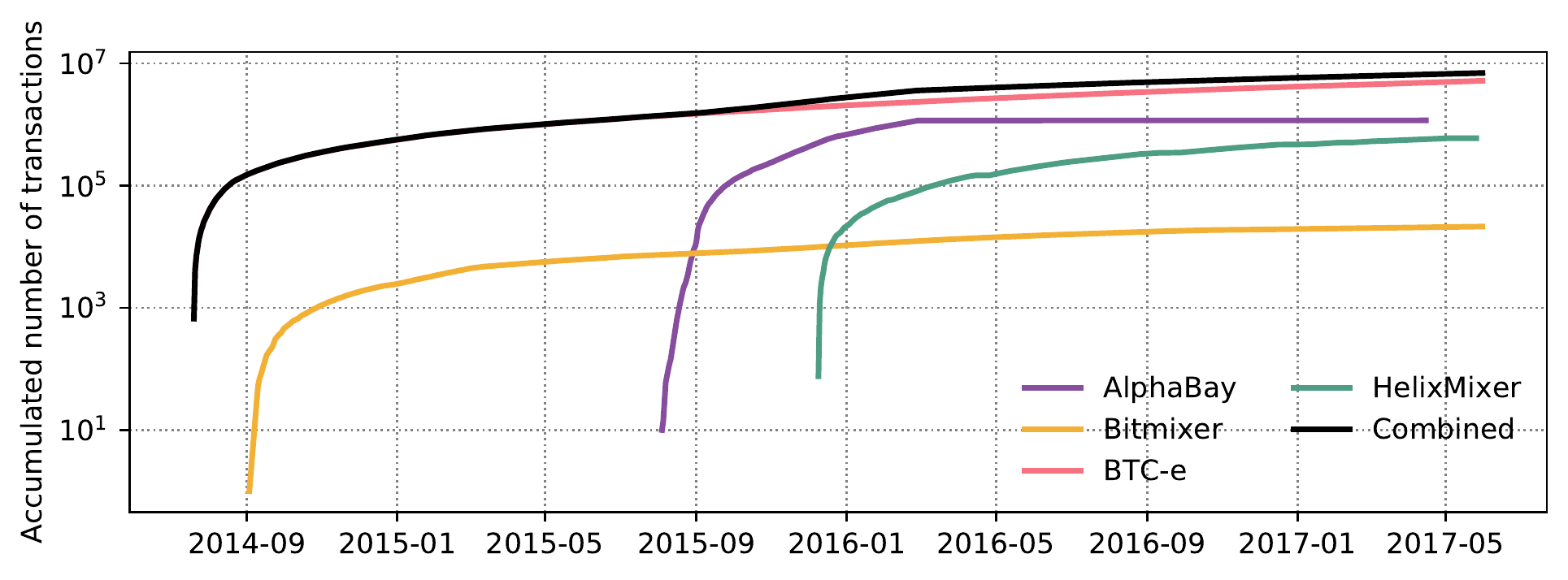} 
	\caption{Service active period and accumulated number of transactions.}
	\label{fig:service_lines}
\end{figure}

\paragraph{Money laundering}
Laundering services disguise the origin of bitcoins by mixing different users' transactions, and many of them are potentially malicious~\cite{moser2013inquiry}
Our selection of \emph{money laundering services} is based on existing literature \cite{moser2013inquiry,de2017analysis}, news articles and address tags from trusted online resources, e.g., WalletExplorer.com~\cite{wallet_explorer}, a website that tracks Bitcoin wallets and aggregates relevant addresses. In total we found 4 major laundering services with more than 22,000 transactions each. Our selection include \emph{AlphaBay}~\cite{AlphaBay}, \emph{BTC-e}~\cite{BTC-e}, \emph{Bitmixer}~\cite{bitmixer} and \emph{HelixMixer}~\cite{de2017analysis}. Figure~\ref{fig:service_lines} shows the accumulated number of daily transactions during the active periods of the four laundering services, as well as a combined sum across all these services. For example, \emph{BTC-e} was active between 08/14-05/17, and generated nearly six million transactions in total.

\vspace{-2mm}
\vspace{-2mm}

\paragraph{Regular}
The exchange sector has the highest number of operating entities, as most new users begin trading on Bitcoin via exchanges~\cite{glaser2014bitcoin}. This is followed by wallet services with $5.8-11.5$ million active users~\cite{rauchs2017global}. Payment services such as \emph{CoinPayment} and \emph{Bitpay} are also often linked to banking institutions and existing payment networks. 
As the lines between different sectors have become increasingly blurred~\cite{rauchs2017global}, when labelling \emph{regular services}, we randomly selected 9 active, large-scale exchange, payment and wallet services from the top wallet list on WalletExplorer.com~\cite{wallet_explorer}, with reference to existing studies~\cite{ranshous2017exchange}. \\

We label all transactions related to regular services as \emph{regular} and those associated with laundering services as \emph{laundering}, similar to the ground truth labelling in \cite{bartoletti2018data}. To ensure the correctness of labelling, we have eliminated all transactions with conflicting labels~\cite{ermilov2017automatic}. Table~\ref{tb:malicious_addr_tx} summarizes our data collection with the total number of laundering and regular transactions, active periods of laundering services, as well as the percentage of labelled laundering and regular transactions among all existing transactions.

\vspace{-2mm}
\begin{table}[!ht]
	\centering
	\caption{Data collection.} 
	\label{tb:malicious_addr_tx}
	\begin{tabular}{m{2.2cm}|m{2.6cm}|m{2.6cm}|m{2cm}}\hline
		\textbf{Service} & \textbf{Number of Tx.s} & \textbf{Active period} & \textbf{Percentage} \\ \hline
		AlphaBay & 1,168,382 & 08/15-04/17 & 0.97\% \\ 
		Bitmixer & 22,122 & 09/14-06/17 & 0.01\% \\ 
		BTC-e & 5,665,400 & 08/14-06/17 & 2.93\% \\ 
		HelixMixer & 605,991 & 12/15-08/16 & 0.56\% \\ \hline 
		Laundering & 7,461,895 & - & 4.29\% \\ 
		Regular & 37,907,769 & - & 22.98\% \\ \hline 
	\end{tabular}
\end{table}
\vspace{-2mm}



%% file: graph_characterization.tex
\section{Graph Characterization}
\label{sec:graph_characterization}
We next provide a characterization of Bitcoin transaction graphs, with a focus on the difference between laundering and regular transactions.

\subsection{Graph Evolution}
\label{subsec:graph_characterization}
\begin{figure*}[!htbp]
	\centering
    	\subfloat[\footnotesize{Node}]{
    		\includegraphics[width=0.325\textwidth]{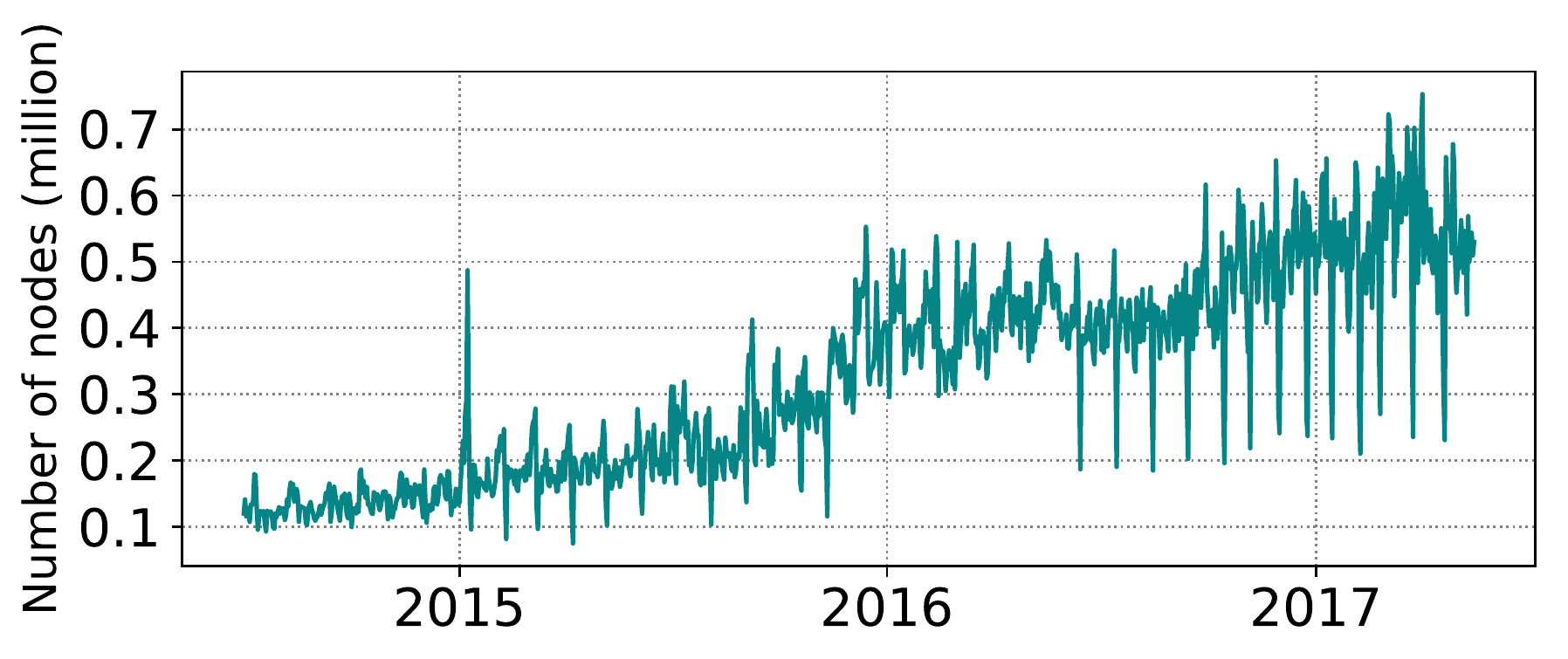} \label{fig:nodes}
    	}
    	\subfloat[\footnotesize{Edge}]{
    		\includegraphics[width=0.325\textwidth]{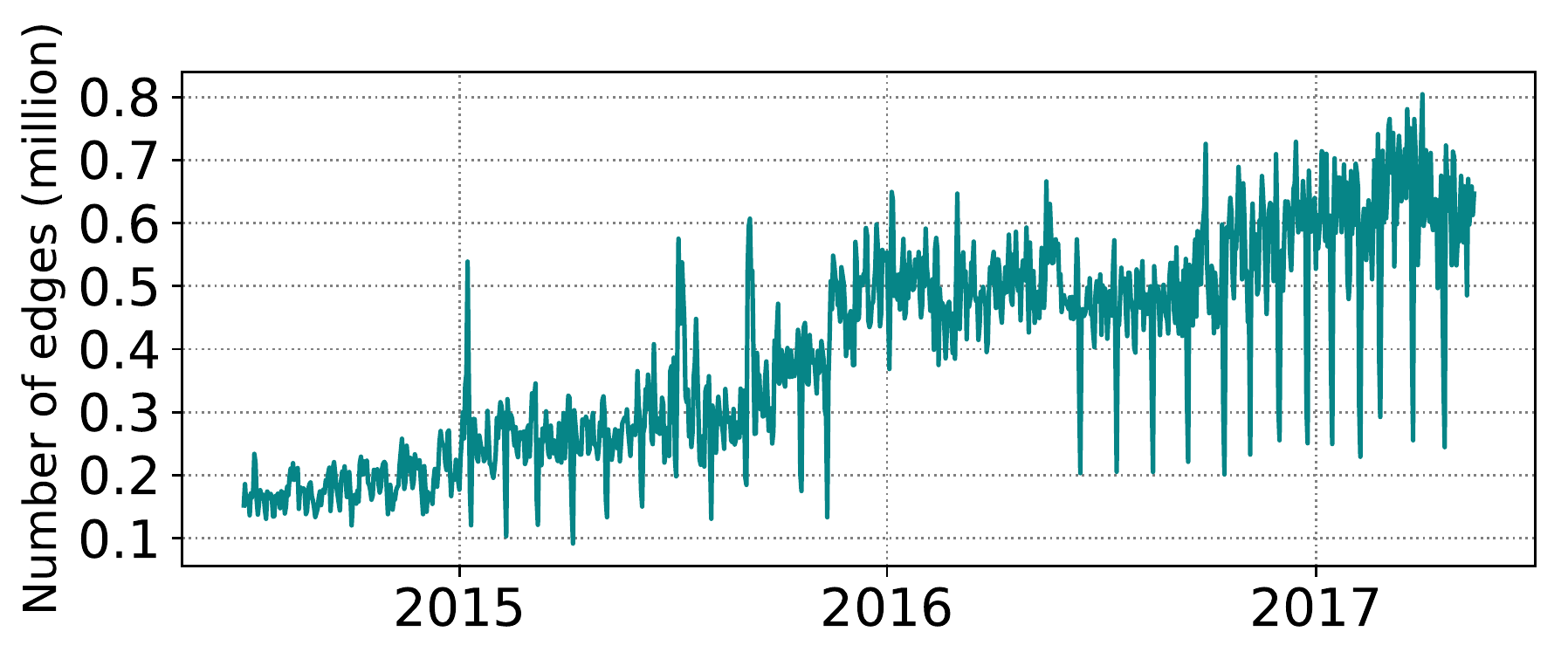} \label{fig:edges}
    	}
    	\subfloat[\footnotesize{Graph density}]{
    		\includegraphics[width=0.325\textwidth]{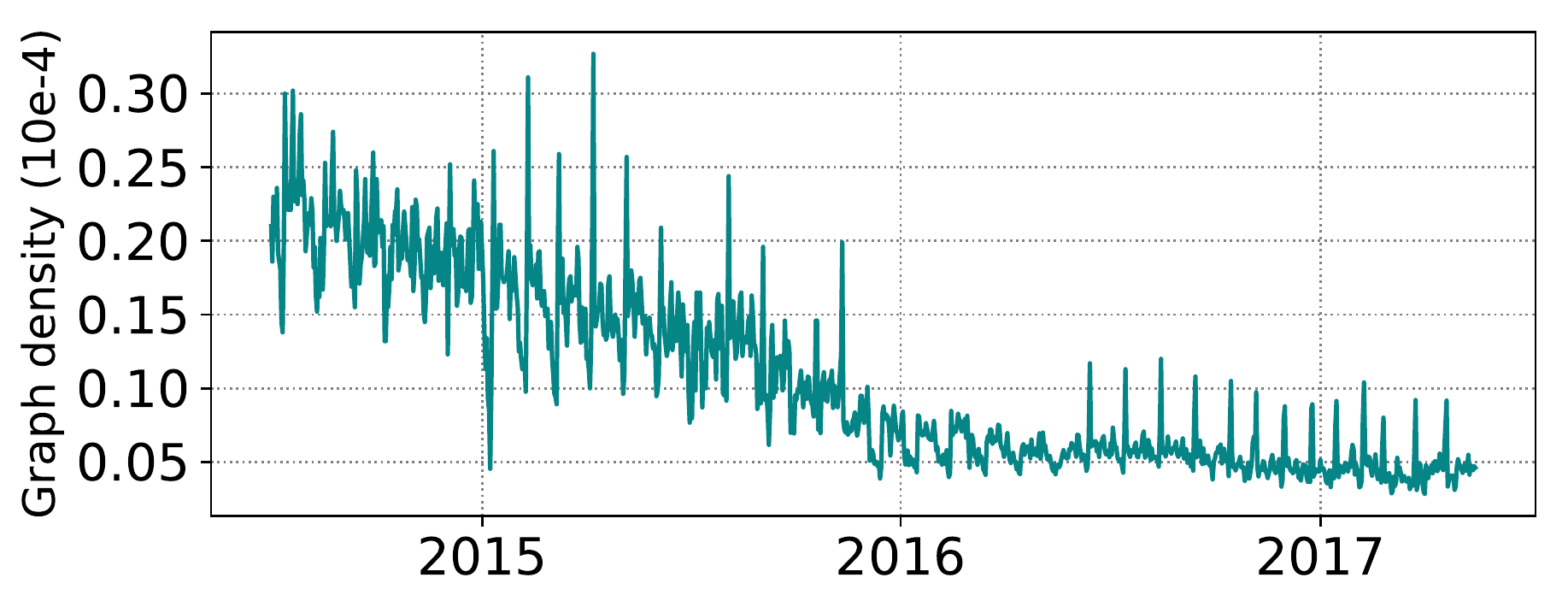} \label{fig:density}
    	}
	\caption{Evolution of daily transaction graphs.} 
	\label{fig:graph_evolution}
\end{figure*}

We first show the growth of Bitcoin network using directed daily transaction graphs created between 07/2014-05/2017. We calculated the number of nodes, number of edges, and graph density for each graph. Figure~\ref{fig:graph_evolution} shows the evolution of these 3 metrics over the approximate 3-year time span. Both number of nodes and number of edges continued to increase, and started to saturate in early 2017. The decrease in graph density suggests transactions have fewer inputs and outputs on average in more recent years. This could be due to the growing number of wallet or payment services that run specific algorithms to construct transactions. 
The periodic fluctuations can be ascribed to the difficulty adjustment embedded in the Bitcoin Proof-of-Work (PoW) protocol~\cite{antonopoulos2014mastering}. As the total hashrate changes, to maintain the constant 10 minute block time, Bitcoin adjusts PoW difficulty every 2016 blocks, which approximately was 2 weeks in 2016~\cite{btc_pow}.

\vspace{-1mm}
\vspace{-1mm}

\subsection{Feature Characterization} \label{subsec:feat_characterization}
\begin{figure}[!htbp]
	\centering
	\includegraphics[width=.85\columnwidth]{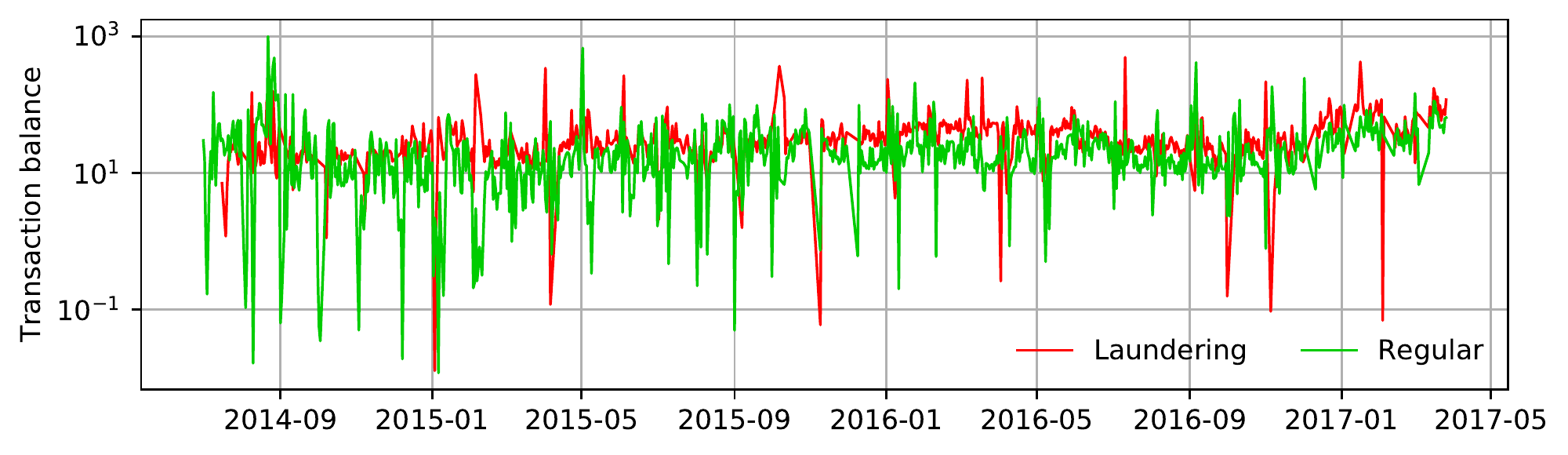}
	\caption{Average daily transaction value by service type.}
	\label{fig:balance_evo}
\end{figure}

We then look at transaction values, calculated as the sum of all output UXTOs. Figure~\ref{fig:balance_evo} shows the daily average value of laundering and regular transactions. 
On average, laundering transactions carry 38.8 bitcoins per day, while regular transactions only have 29.1 bitcoins per day. Laundering transactions also carry higher values than regular ones 77.3\% of the time.

\vspace{-1mm}
\vspace{-1mm}

For each directed daily transaction graph, we calculate basic features such as number of input UTXOs (or in-degree), number of output UTXOs (or out-degree), sum/mean/standard deviation (std) of outputs, and network features such as centrality measures, PageRank, and clustering co-efficient~\cite{bondy1976graph}. We extracted a total of 14 features for each transaction. We performed a feature importance analysis using an \emph{Adaptive Boost (Adaboost)}~\cite{freund1996experiments} classifier with a Decision Tree based estimator of maximum depth 5. We selected 5 most discriminating features based on the importance score, among which the in-degree/out-degree ratio provides information on edges; the sum/mean/std of outputs relate to UTXO values; and the weakly connected components relate to a transaction's neighbourhood.

\begin{figure}[!htbp]
	\centering
	\resizebox{\textwidth}{!}{%
    	\subfloat[\footnotesize{In/Out Ratio}]{%
    		\includegraphics[width=0.22\textwidth]{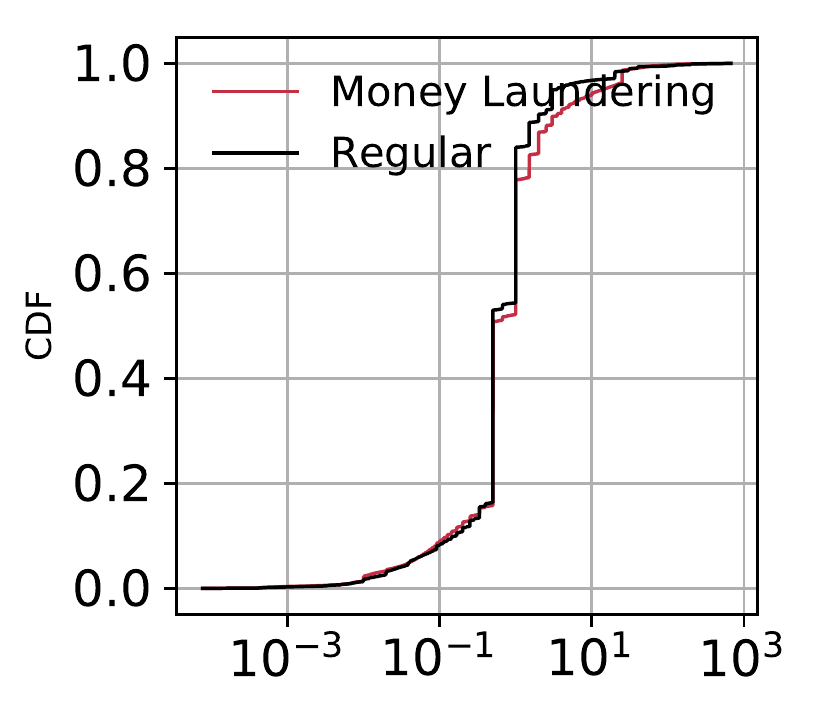}
    		\label{fig:ml_ratio}
    	}
    	\subfloat[\footnotesize{Sum of outputs}]{%
    		\includegraphics[width=0.22\textwidth]{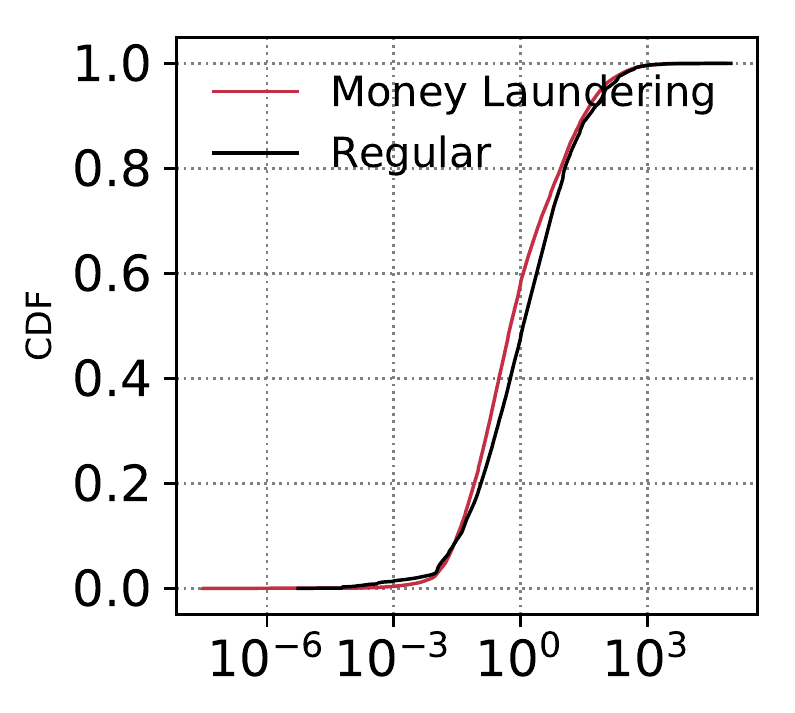}
    		\label{fig:ml_sum}
    	}
    	\subfloat[\footnotesize{Mean of outputs}]{%
    		\includegraphics[width=0.22\textwidth]{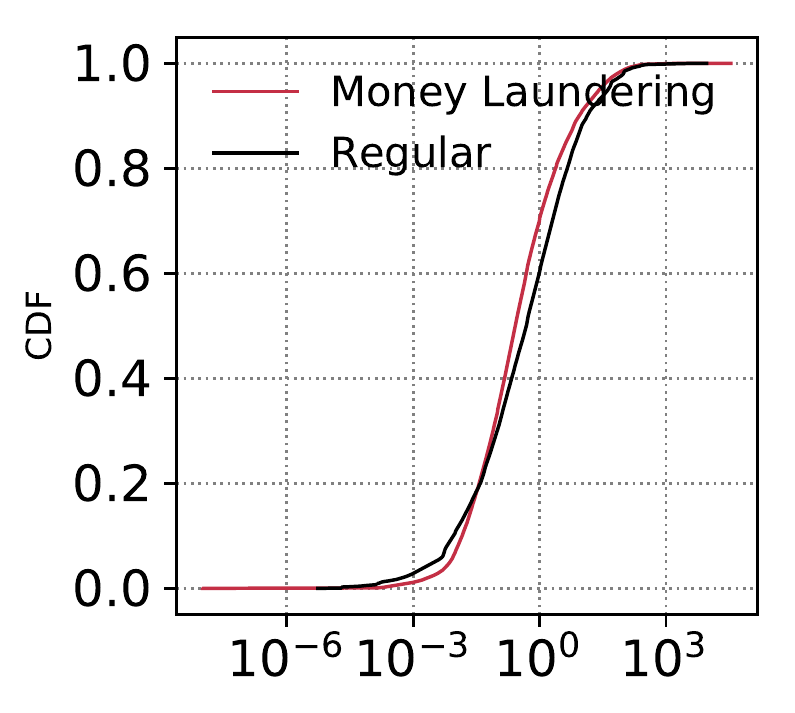}
    		\label{fig:ml_mean}
    	}
    	\subfloat[\footnotesize{Std of outputs}]{%
    		\includegraphics[width=0.22\textwidth]{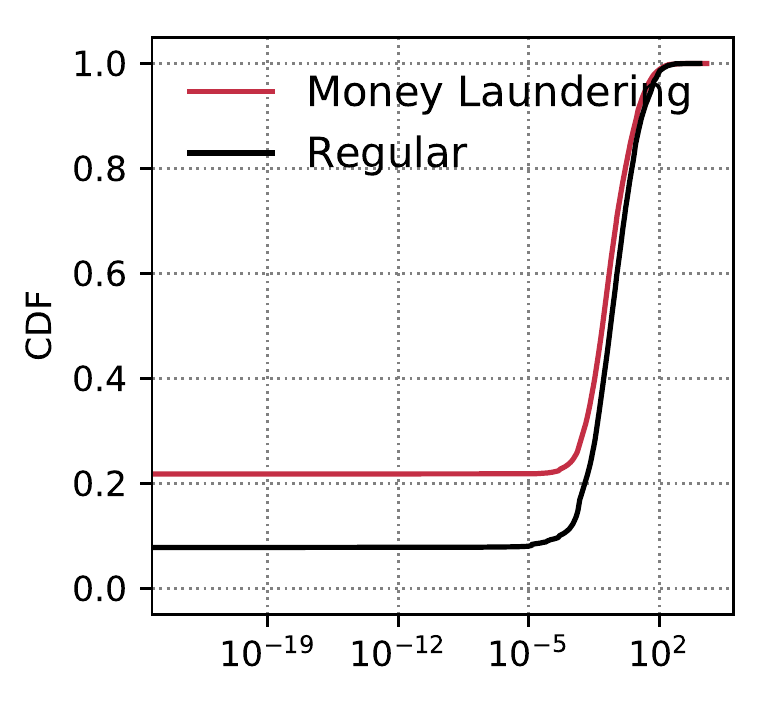}
    		\label{fig:ml_std}
    	}
    	\subfloat[\footnotesize{Components}]{%
    		\includegraphics[width=0.22\textwidth]{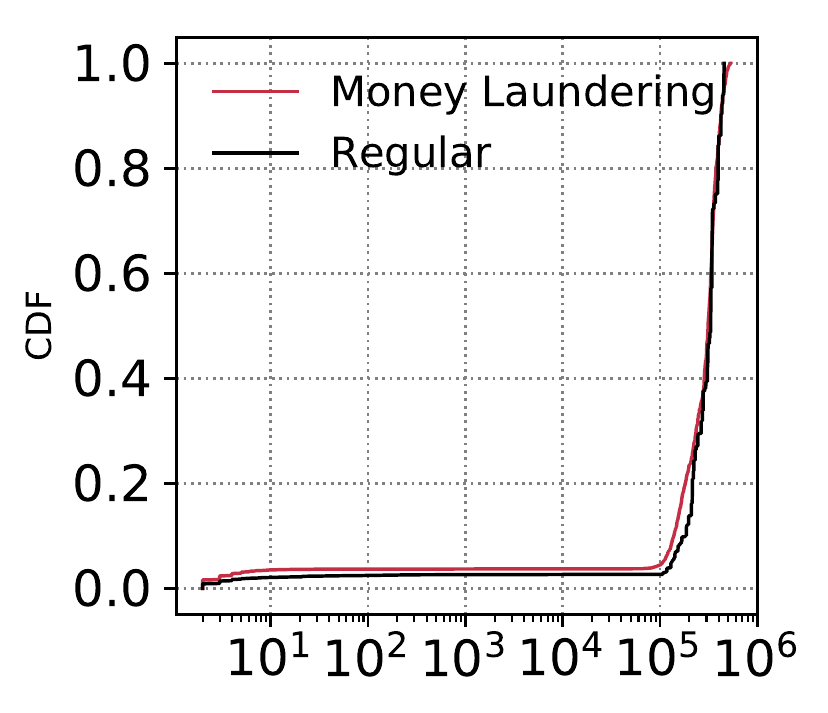}
    		\label{fig:ml_weak}
    	}
    }
	\caption{Feature distribution for money laundering and regular transactions.}
	\label{fig:mixing_regular_feats}
\end{figure}

Figure~\ref{fig:mixing_regular_feats} shows the distribution of these 5 features of laundering and regular transactions separately. Compared to regular transactions, a larger portion of laundering transactions have high in-degree/out-degree ratio. The resulting distribution confirms laundering services operate by bundling small transactions from various users and forming new ones, which is also discussed in Section~\ref{sec:related}.
The CDF curves for sum, mean and std of output UTXOs of laundering transactions are slightly steeper than those of regular transactions. This indicates laundering transactions create similar number and value of output UTXOs, which also confirms the findings in~\cite{moser2013inquiry}.
Laundering transactions also have a slightly smaller number of weakly connected components than regular transactions. This is resulted from the scales of different services. As services tend to send and receive transactions using addresses associated to them, directly connected transactions are more likely to belong to a same service. We leverage this observation later to design the baseline classifier in Section~\ref{sec:classification}.


\vspace{-2mm}
\vspace{-2mm}

\begin{figure}[!htbp]
	\centering
	\resizebox{\textwidth}{!}{%
    	\subfloat[\footnotesize{Ratio}]{%
    		\includegraphics[width=0.22\textwidth]{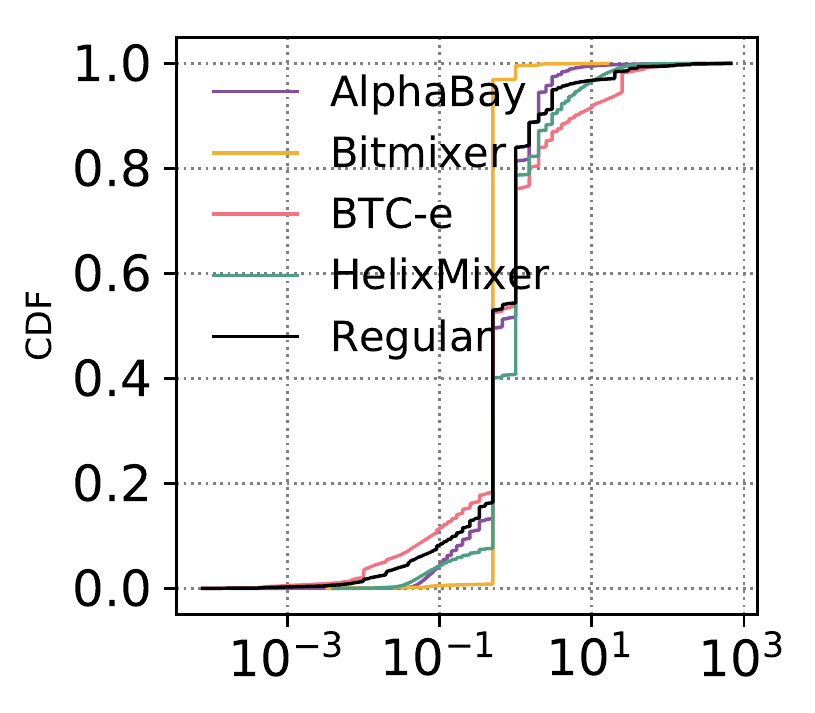}
    		\label{fig:comp_ratio}
    	}
    	\subfloat[\footnotesize{Sum of outputs}]{%
    		\includegraphics[width=0.22\textwidth]{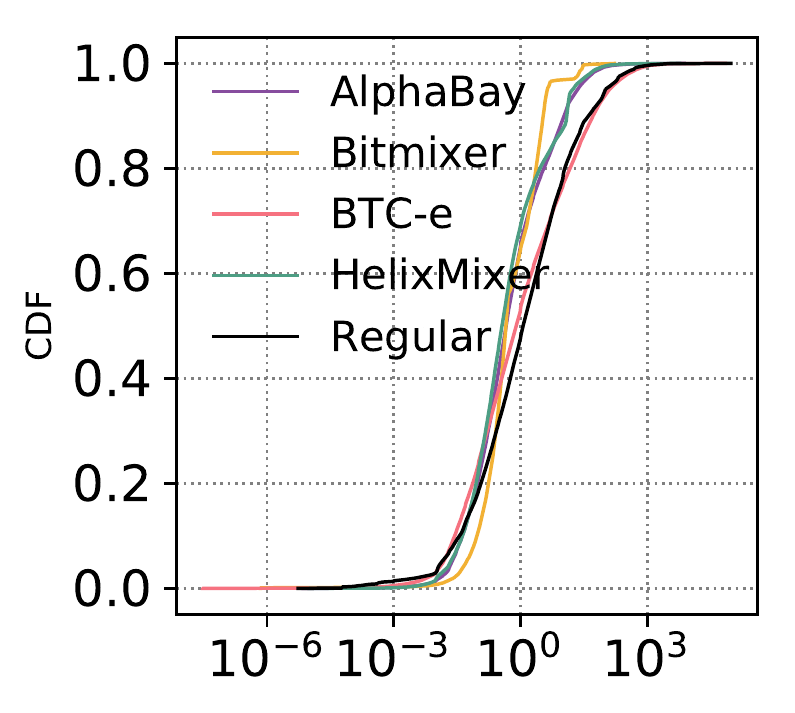}
    		\label{fig:comp_sum}
    	}
    	\subfloat[\footnotesize{Mean of outputs}]{%
    		\includegraphics[width=0.22\textwidth]{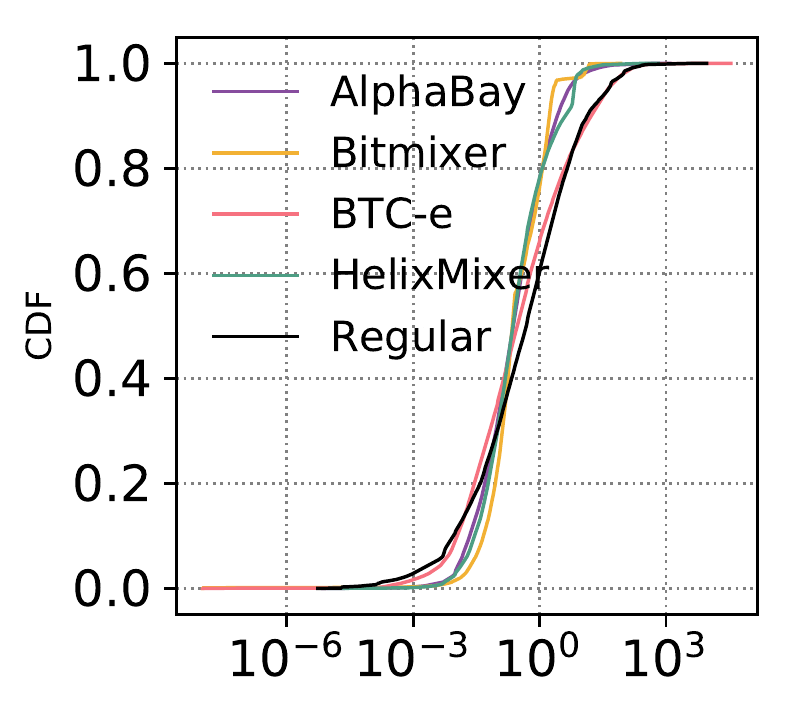}
    		\label{fig:comp_mean}
    	}
    	\subfloat[\footnotesize{Std of outputs}]{%
    		\includegraphics[width=0.22\textwidth]{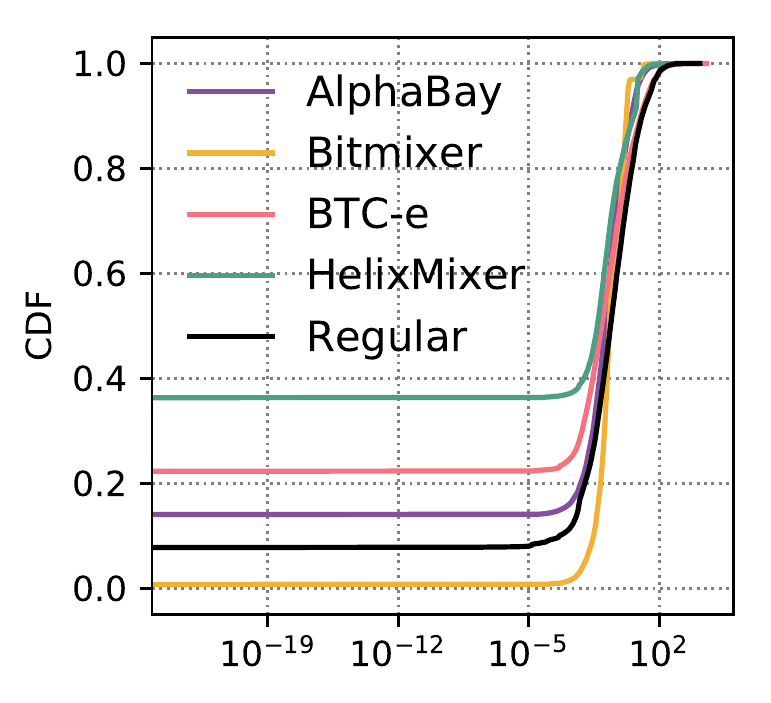}
    		\label{fig:comp_std}
    	}
    	\subfloat[\footnotesize{Components}]{%
    		\includegraphics[width=0.22\textwidth]{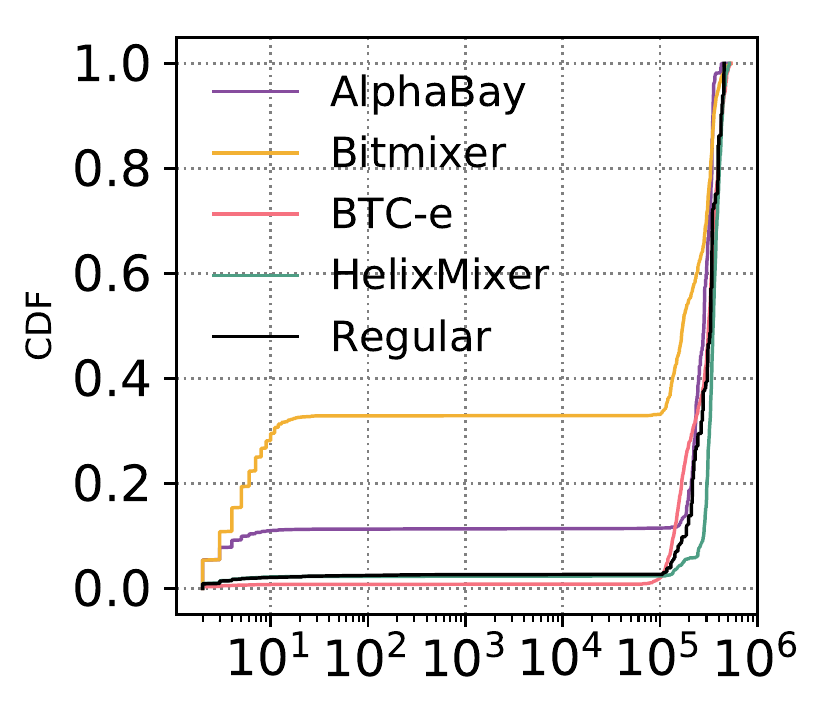}
    		\label{fig:comp_weak}
    	}
    }
	\caption{Service-wise feature distribution.} 
	\label{fig:service_feats}
\end{figure}

To further understand the service-wise differences, we provide a more detailed comparison in Figure~\ref{fig:service_feats}. Although there are some differences among laundering transactions related to different services, these transactions are individually distinguishable from regular ones in the selected features. Among the 4 laundering services studied, \emph{BTC-e} shows the most similar feature distributions as regular transactions. \emph{Bitmixer}, on the other hand, behaves similarly to \emph{HelixMixer}.


\vspace{-2mm}

%% file: service_combined.tex
\section{Detecting Money Laundering Activities}
\label{sec:classification}

We next build multiple classifiers to detect laundering transactions using the partially labelled dataset described in Section~\ref{sec:data}. We train our classifiers with four different features--nodes' immediate neighbours, curated features, deepwalk embeddings and node2vec embeddings. 
We used accuracy--the total percentage of regular and laundering transactions that are correctly identified, and F1-measure--the harmonic mean of precision and recall--as evaluation metrics.

\subsection{Transaction Classifiers}
The design details of our four classifiers are discussed below.

\paragraph{\bf Immediate Neighbour-Based Classifier}
Since transactions from a same service tend to be more connected (cf. Section~\ref{subsec:feat_characterization}), as a first step, we classify unlabelled transactions based on their nearest neighbours. This classifier works under the following criteria: 1) a node with more immediate regular neighbours than laundering neighbours is classified regular, 2) a node with more or equal number of immediate laundering than regular neighbours is classified laundering, and 3) a node with no regular or laundering neighbours is classified regular--this criterion is required due to the unlabelled nodes in the graph.

\paragraph{\bf Curated Features}
We trained an Adaboost classifier with a Decision Tree base-estimator using the 14 statistical and network features discussed in Section~\ref{sec:graph_characterization}. Hyper-parameter tuning results when varying the maximum depth and number of estimators are presented in Appendix~\ref{app:para_tune}.

\paragraph{\bf Deepwalk}
Graph representation learning via random walks has become a common technique to analyse graph structures in recent years~\cite{hamilton2017representation}. These algorithms can automatically create feature vectors for graph nodes in an unsupervised manner, and achieve better scalability and performance than manually extracted features in various scenarios. A widely used algorithm--\emph{deepwalk}~\cite{perozzi2014deepwalk}--leverages random walks of specified \emph{number of walks per node} and \emph{walk length} to uniformly sample a node's neighbourhood.

\paragraph{\bf Node2vec}
A more recent technique, \emph{node2vec}~\cite{grover2016node2vec}, uses 2 more parameters--the return parameter \emph{p} and the in-out parameter \emph{q}--to more precisely guide the walks. 
When \emph{p} is high ($>max(q,1)$), the same nodes will not be revisited in the next 2 steps~\cite{grover2016node2vec}, and when \emph{p} is low ($<max(q,1)$), the walks will remain close to the starting node. When \emph{q} and \emph{p} both equal to 1, node2vec creates random walks in a uniform manner, similar to deepwalk. \\

Deepwalk and node2vec are strictly transductive~\cite{hamilton2017inductive}, and can only predict unlabelled nodes on the same graph, hence we created transaction graphs that cover the entire duration of training and testing, and only considered nodes on the giant component~\cite{bollobas2001evolution}. We also used undirected graphs to better explore the graph structure. 
For both binary classification and new instance prediction in Section~\ref{sec:prediction}, we started random-walks from all labelled nodes in the training set and all labelled and unlabelled nodes in the testing set to ensure no pre-assumed knowledge on the testing nodes. 
We used 100 (\emph{number of walks per node}), 100 (\emph{walk length}), 25 (\emph{embedding dimension}) for both deepwalk and node2vec, and 2 (\emph{p}), and 0.5 (\emph{q}) for node2vec.
We fed the embedding vectors into an Adaboost classifier with a Decision Tree base classifier (\emph{maximum depth} = 5, \emph{number of estimators} = 40) for binary classification. Details on the selection of random-walk parameters and walk starting nodes are provided in Appendix~\ref{app:para_tune}.

\subsection{Classifier Performance Comparison}
We set the experiment duration to 1 week with the first 5 days for training and last 2 days for testing. To compare the performance of the above 4 classifiers, we selected 5 consecutive weeks in late 2014, and present the results averaged over these weeks below. 
In addition, we also applied two simple ensemble techniques to the prediction results using \emph{curated features} and \emph{node2vec embeddings}. In ``OR" ensemble, we label a transaction as laundering if either classifier labels it as laundering. While in ``AND" ensemble, we only label a transaction as laundering if both classifiers do so.

Table~\ref{tb:classifier_performance} shows the node2vec-based classifier with proper parameter settings outperforms the other three classifiers achieving an average accuracy of 92.05\% and an average F1-measure of 0.94.
Information from a transaction's immediate neighbours is not sufficient in detecting laundering transactions. Although some differences can be observed in the feature distribution (cf. Section~\ref{sec:graph_characterization}), manually extracted statistical and network features cannot effectively differentiate laundering and regular transactions. Calculating network features on these large transaction graphs can also be very time-consuming. 
Both deepwalk and node2vec performed well in binary classification, while node2vec produced slightly better results than deepwalk.
``OR" ensemble improved the node2vec classifier performance as it helps to capture more laundering nodes. ``AND" ensemble on the other hand drastically diminished the performance of both classifiers, as more laundering transactions are wrongly detected as regular.

\vspace{-2mm}
\begin{table}[!ht]
	\centering
	\caption{Classifier Performance.} 
	\label{tb:classifier_performance}
	\begin{tabular}{m{4.4cm}|m{2.6cm}|m{2.2cm}}\hline
		\textbf{Classifier} & \textbf{Accuracy (\%)} & \textbf{F1-measure} \\ \hline
		Neighbourhood & 28.46 & 0.09 \\
		Manually extracted features & 65.34 & 0.45 \\
        Deepwalk & 91.72 & 0.94 \\
		Node2vec & 92.05 & 0.94 \\ \hline 
		``OR" ensemble & 92.74 & 0.95 \\
		``AND" ensemble & 21.47 & 0.02 \\ \hline
	\end{tabular}
\end{table}
\vspace{-2mm}

\subsection{Over-time Classification Results With Node2vec}

\begin{figure}[!htbp]
	\centering
	\includegraphics[width=.9\columnwidth]{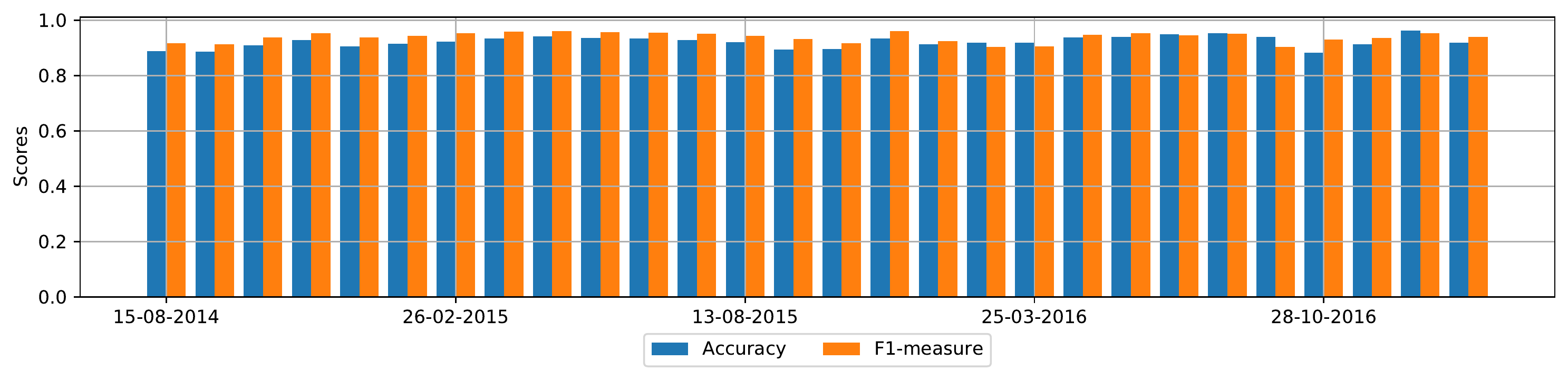}
	\caption{Node2vec classifier performance over time.}
	\label{fig:full_classification}
\end{figure}

We then applied the node2vec-based classifier with proper parameters on one random week per month between 08/2014-01/2017.
Figure~\ref{fig:full_classification} illustrates the classifier performance over the entire experimentation period. To ensure training reliability, we have removed scenarios with less than 150 laundering samples in training. The classifier performance was robust and achieved an average accuracy of 92.29\% and F1-measure of 0.93 across the selected weeks.

\vspace{-1mm}
\vspace{-1mm}



%% file: service_separate.tex
\section{Predicting New Money Laundering Instances} \label{sec:prediction}
\begin{figure}[!htbp]
	\centering
	\subfloat[\footnotesize{Accuracy}]{%
		\includegraphics[width=0.5\textwidth]{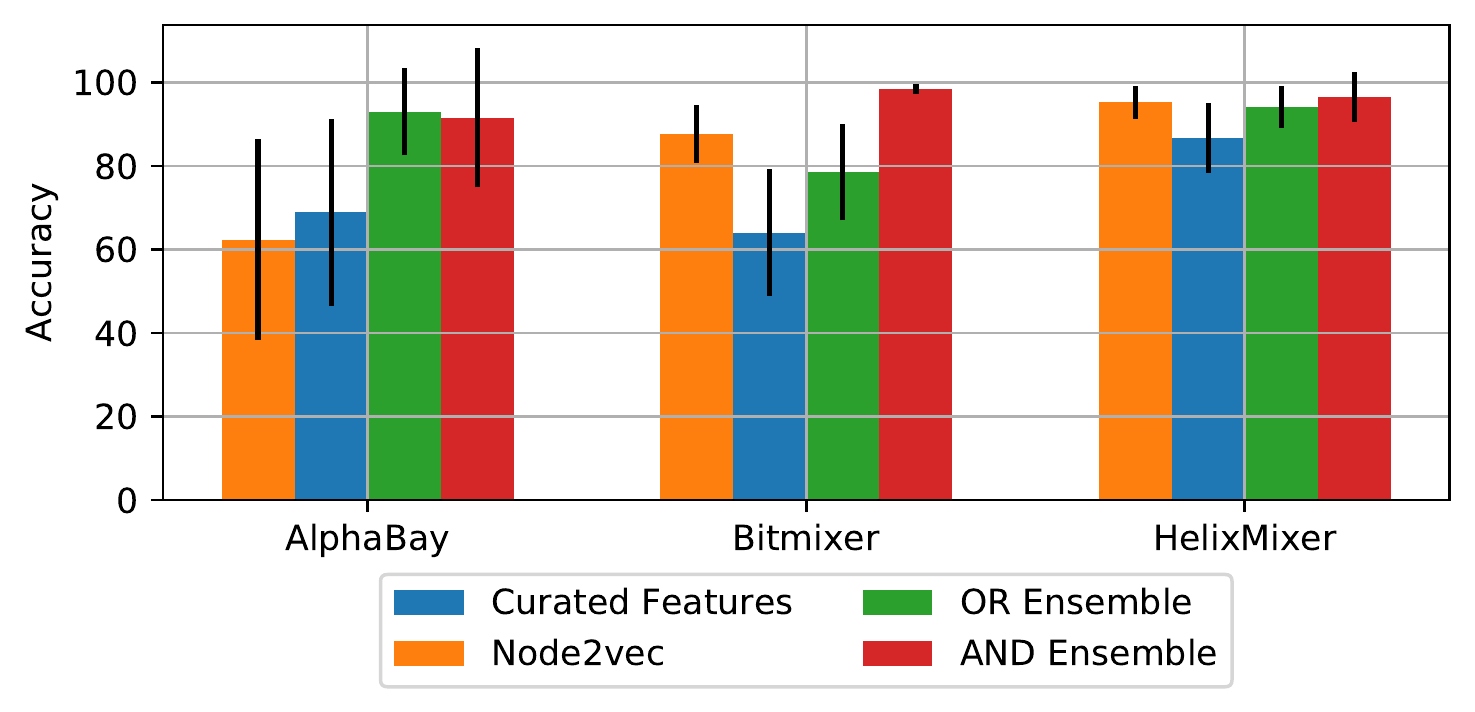}
		\label{fig:service_acc}
	}
	\subfloat[\footnotesize{F1-measure}]{%
		\includegraphics[width=0.5\textwidth]{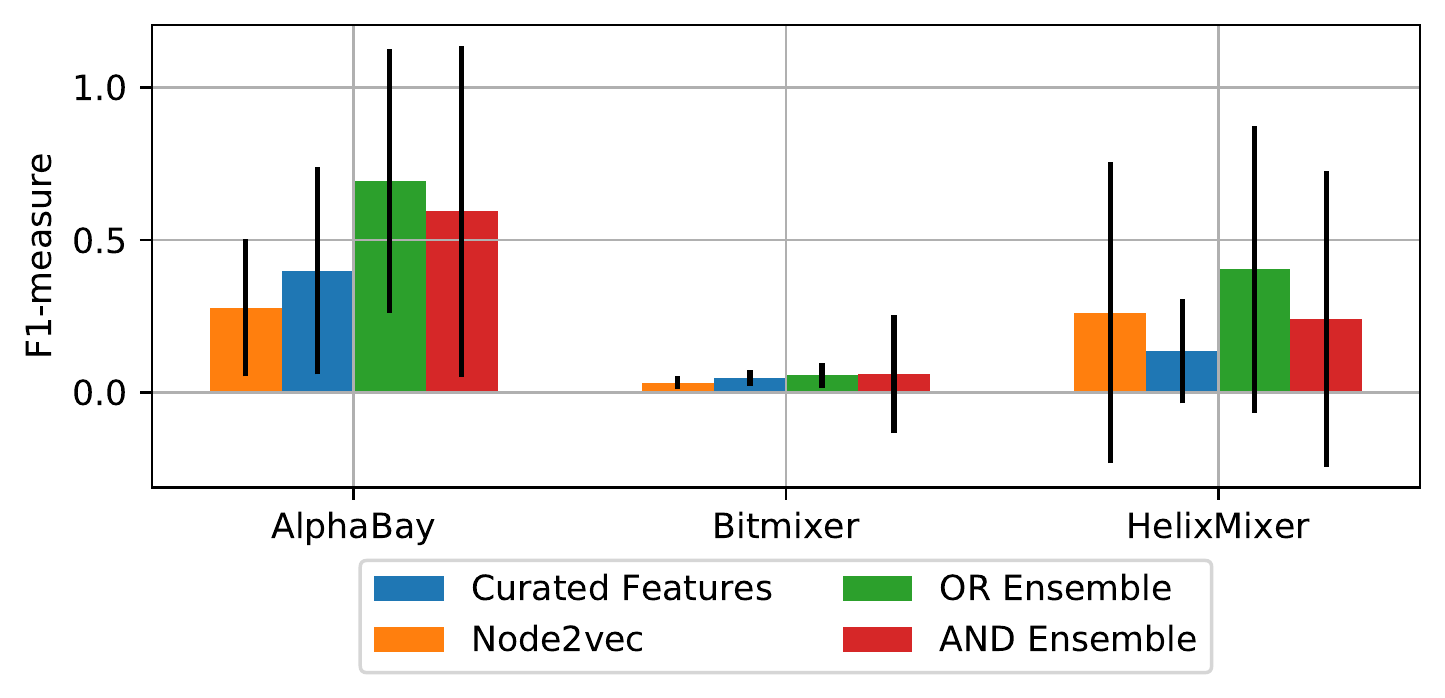}
		\label{fig:service_fmeas}
	}
	\caption{Comparison of classifier performance for different services.} 
	\label{fig:new_service_comparison}
\end{figure}

We next apply a curated feature-based classifier and a node2vec-based classifier with corresponding parameters to detect previously unseen laundering transactions. We selected \emph{AlphaBay}, \emph{Bitmixer} and \emph{HelixMixer} for comparison. For every 2 or 3 months during a service's active period, we randomly selected one week to leave the target service out and train the model using the remaining services. Then, we evaluated the model's ability to discover the left-out service. 
Figure~\ref{fig:new_service_comparison} compares the classifier performance and prediction ensemble for the 3 selected services, averaged over their respective experiment periods.

Compared to the classification results in Section~\ref{sec:classification}, the classifier performance decreases when predicting instances from new services.
Nonetheless, the node2vec classifier was able to achieve 95.2\% accuracy in detecting \emph{HelixMixer} with a F1-measure of 0.3.
As observed in Figure~\ref{fig:service_feats}, there are differences in several features among these services.
When applying node2vec, the random walks cannot effectively explore the neighbourhoods of transactions belonging to a different service. The class imbalance also becomes more significant when splitting training and testing sets by service, resulting in more malicious nodes being classified as regular. ``OR" ensemble improves the F1-measure, as it tends to capture more malicious transactions. ``AND" ensemble on the other hand improves the accuracy as more regular transactions are correctly identified. 

In the case of \emph{Bitmixer}, there are on average 66 laundering transactions and 24,428 regular transactions in the test set of each week. 
This significant class imbalance causes an extremely low F1-measure, which is not improved even with the proposed ensemble techniques.
As for \emph{AlphaBay} and \emph{HelixMixer}, which have discriminating features compared to regular transactions, the F1-measure using the curated feature-based classifier increases as more laundering transactions can be easily distinguished from regular ones. Both services also have sufficient samples, with an average money laundering to regular ratio being 10,485/18,727 and 2,287/24,842 respectively. When applying node2vec, their transactions are more often covered by random walks and are hence more correctly identified.

Since the prediction results are also affected by the number of testing samples, it can be inferred that laundering services need to grow to a certain extent for our classifiers to be effective. Services with a small number of transactions are hard to predict, and hence they can operate unnoticed and last longer than services who create a higher transaction volume.

\vspace{-1mm}
\vspace{-1mm}

%% file: discussion_conclusion.tex
\section{Concluding Remarks} 
\label{sec:discussion_conclusion}
We characterized Bitcoin transaction graphs and studied patterns of money laundering and regular transactions on data collected over three years. 
Our study found that money laundering transactions tend to have a higher in-degree/out-degree ratio, more uniform sum, mean and std of outputs, and a slightly smaller number of weakly connected components compared to regular transactions. We also show that \emph{BTC-e} behaves similarly to regular services, while \emph{Bitmixer} and \emph{HelixMixer} have the most similar behaviors among the 4 selected laundering services.
The preliminary classification and prediction results using an Adaboost classifier proved we can differentiate money laundering from regular transactions, and predict new instances using node2vec embeddings. Results also showed the classifier performance can be improved with ensemble techniques. Future work will mainly focus on the two aspects below.

\paragraph{Ground Truth Labelling}
We operated on a fraction ($\approx 27\%$) of Bitcoin transactions belonging to a limited number of identified services. A larger fraction of transactions remain unlabelled. Finding sufficient volumes of reliably labelled data for potentially illegal activities such as money laundering is always challenging. As we used data labelled by volunteers, there exist conflicting labels for certain transactions. In this paper, we limit ourselves to the most reliable tags based on information from existing literature and trusted websites. In the future, semi-supervised approaches such as \emph{label propagation}~\cite{zhu2002learning} and \emph{label spreading}~\cite{zhou2004learning} can be leveraged to further improve the performance.

\paragraph{Transductive Graph Embeddings}
As mentioned in Section~\ref{sec:classification}, node2vec is strictly transductive and can only be used to predict unknown nodes on the same graph. One requirement for analysing temporal graphs is the possibility of making predictions to previously unseen nodes. Recent developments based on graph convolutions such as GraphSage~\cite{hamilton2017inductive} which allow embeddings learned from one graph to be used in prediction on a completely different graph can be explored. Nonetheless, their success on very large and dynamic graphs remains to be seen.

%% file: appendix.tex
\newpage
\appendix
\section{Parameter Tuning} \label{app:para_tune}

\subsection{Curated Features}
\begin{figure*}[!htbp]
	\centering
	\subfloat[\footnotesize{Accuracy heatmap}]{%
		\includegraphics[width=0.5\textwidth]{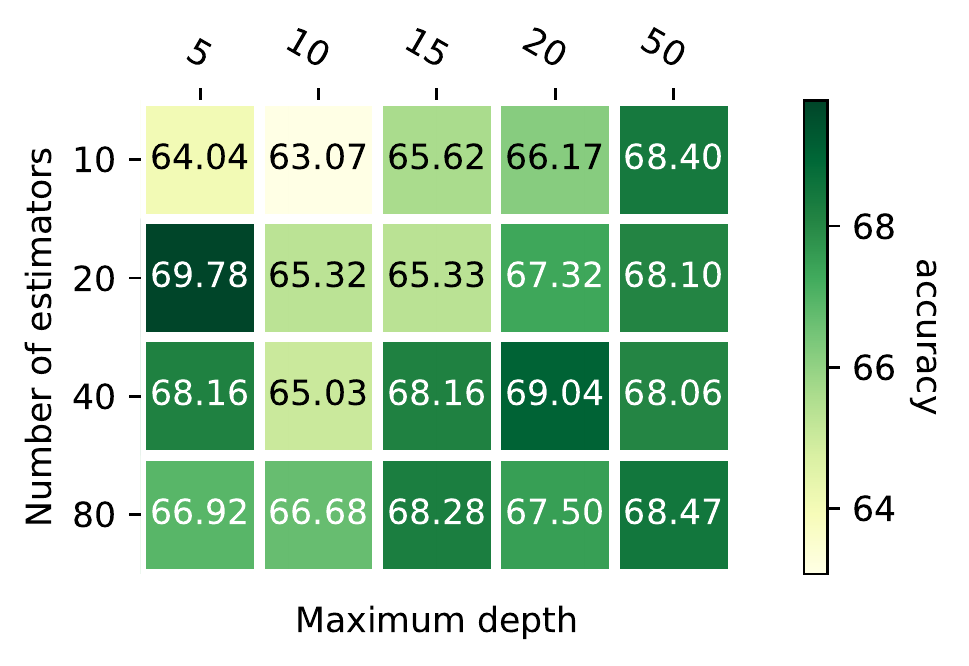}
		\label{fig:para_acc_heatmap}
	}
	\subfloat[\footnotesize{F1-measure heatmap}]{%
		\includegraphics[width=0.5\textwidth]{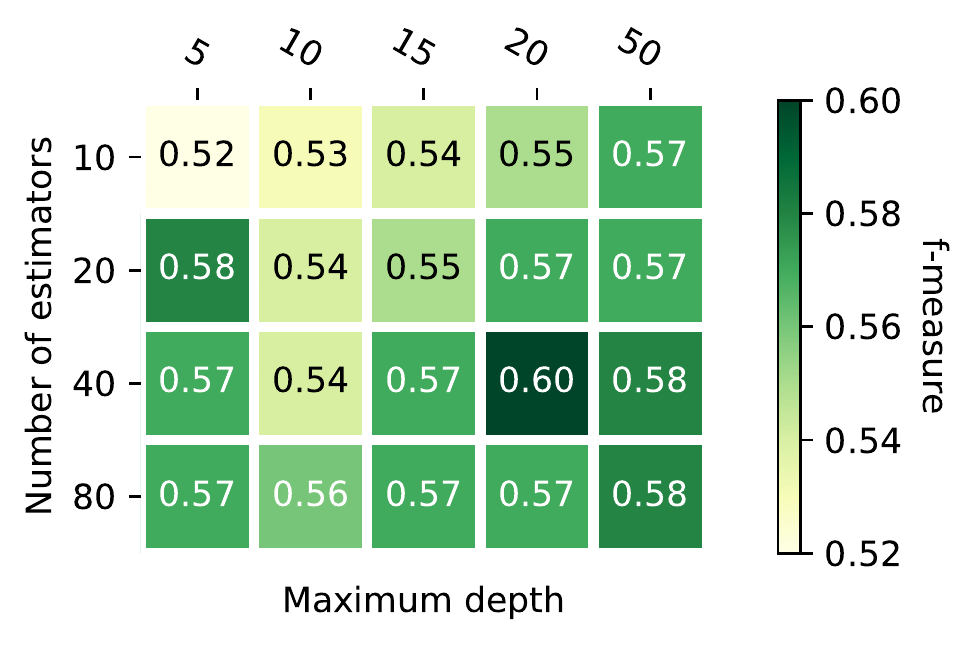}
		\label{fig:para_fmeas_heatmap}
	}
	\caption{Adaboost classifier parameter tuning.} 
	\label{fig:adaboost_paratune}
\end{figure*}

We varied the maximum depth of the base estimator from 5 to 50 and number of estimators from 10 to 80 as illustrated in Figure~\ref{fig:adaboost_paratune}. The classifier performance remains low and does not vary significantly with maximum depth and number of estimators. This confirms manually-extracted statistical and network features are not effective in distinguishing laundering and regular transactions.

\subsection{Node2vec/Deepwalk}
\paragraph{Hyper-parameter Tuning}
Results in Table~\ref{tb:node2vec_paraTune} show that a higher number of walks per node and longer walk path produce better performance. Also, for node2vec it is better to set \emph{q} to a relatively low value and \emph{p} higher than $max(q,1)$ to ensure outwards exploration from transaction nodes, as it is important to understand how UTXOs are propagated via transactions for different services.

\begin{table}[!ht]
\centering
\caption{Node2vec/Deepwalk hyper-parameter tuning.}\label{tb:node2vec_paraTune}
\begin{threeparttable}
\begin{tabular}{m{3.4cm}|m{2cm}|m{0.7cm}|m{0.7cm}|m{2.4cm}|m{2cm}}\hline
	\textbf{\# of walks per node} & \textbf{Walk length} & \textbf{\emph{p}} & \textbf{\emph{q}} & \textbf{Accuracy (\%)} & \textbf{F1-measure} \\ \hline
	50 & 50 & 1 & 1 & 90.20 & 0.93 \\
	50 & 50 & 2 & 0.5 & 90.48 & 0.93 \\
	50 & 50 & 4 & 0.5 & 89.72 & 0.92 \\
	50 & 50 & 0.5 & 2 & 88.35 & 0.91 \\
	50 & 50 & 0.5 & 4 & 86.85 & 0.90 \\ \hline
	100 & 100 & 1 & 1 & 89.84 & 0.92 \\
	100 & 100 & 2 & 0.5 & 91.01 & 0.93 \\
	100 & 100 & 4 & 0.5 & 89.62 & 0.92 \\ \hline
\end{tabular}
\begin{tablenotes}
    \item[*] The graph has 634,739 nodes and 1,147,979 edges. 
\end{tablenotes}
\end{threeparttable}
\end{table}

\begin{figure*}[!htbp]
	\centering
	\subfloat[\footnotesize{Accuracy heatmap}]{%
		\includegraphics[width=0.5\textwidth]{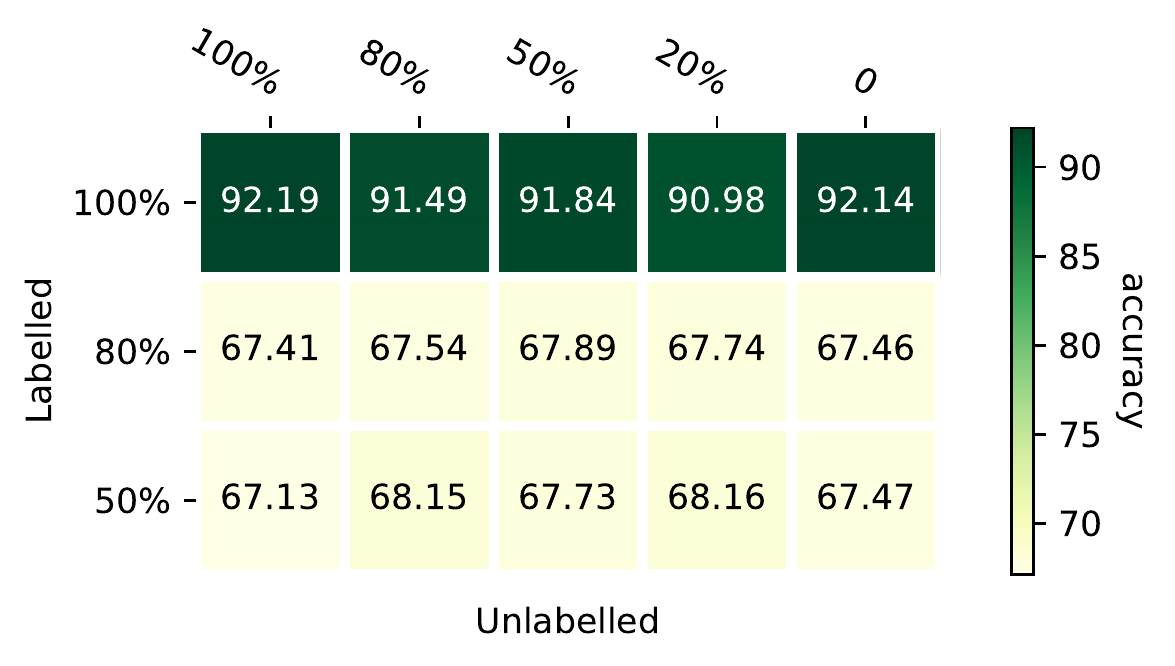}
		\label{fig:walk_acc_heatmap}
	}
	\subfloat[\footnotesize{F1-measure heatmap}]{%
		\includegraphics[width=0.5\textwidth]{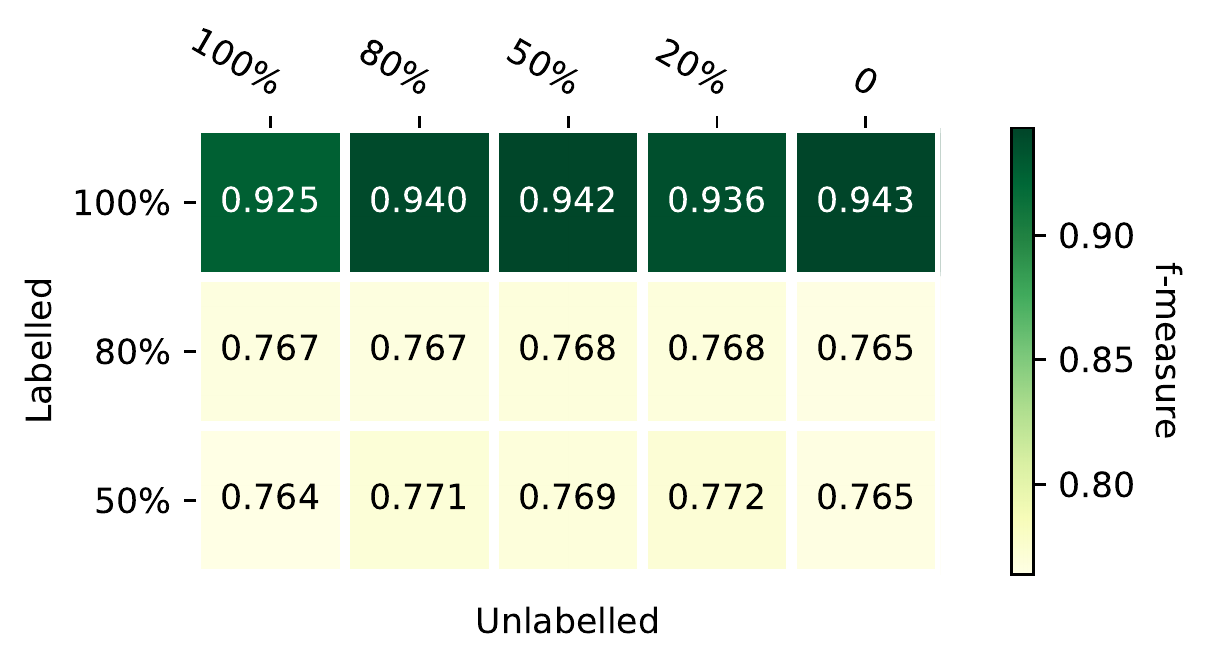}
		\label{fig:walk_fmeas_heatmap}
	}
	\caption{Effect of random-walk starting nodes on classifier performance.} 
	\label{fig:heatmaps}
\end{figure*}

\paragraph{Effect of Random-Walk Starting Nodes} 
Further, to understand the effect of random-walk starting nodes, we compared the classifier performance when varying the volume of labelled and unlabelled test nodes to start random-walks from and present the results in Figure~\ref{fig:heatmaps}. The random-walks are started from all labelled training nodes, x\% ($x \in \{100,80\}$) of all labelled testing nodes, y\% ($y \in \{100,80,50,20,0\}$) of all unlabelled testing nodes. Results show that to achieve high classifier performance, random-walks should be started from as many labelled testing nodes as possible. The volume of unlabelled random-walk starting nodes does not significantly affect the classifier performance. 

